# SUPERSTRIPES

*Self organization of quantum wires in high $T_c$ superconductors*

A. Bianconi, D. Di Castro, N. L. Saini
*Unità INFM, Dipartimento di Fisica, Università di Roma La Sapienza,*
*00185 Roma, Italy*

G. Bianconi,
*Department of Physics, Notre Dame University, 46556 NotreDame,*
*Indiana*

Abstract   We report experimental evidence for the phase diagram of doped cuprate superconductors as a function of the micro-strain ε of the Cu-O bond length, measured by Cu K-edge EXAFS, and hole doping δ. This phase diagram shows a QCP at $P(\varepsilon_c,\delta_c)$ where for $\varepsilon>\varepsilon_c$ charge-orbital-spin stripes and free carriers co-exist. The superconducting phase occurs in the region of critical fluctuations around this QCP. The function $T_c(\varepsilon,\delta)$ of two variables shows its maximum at the strain driven QCP. The critical fluctuations near this strain QCP give the self-organization of a metallic superlattice of quantum wires "superstripes" that favors the amplification of the critical temperature.

## 1. INTRODUCTION

High $T_c$ cuprate perovskites provide an exotic superconducting phase at half way between absolute zero temperature and room temperature. Conventional superconductivity appears in metals with a very high charge density ρ and near absolute zero temperature. In these materials the electrons in the normal phase, above the critical temperature $T_c$ can be considered as free particles following the Fermi statistics (fermions) being in the high-density limit and at low temperature. The electrons are described by a single particle wavefunction Ψ(r) that gives the probability $\Psi^2(r)$ to find an electron in the point r. Below $T_c$ electron pairs condense into a single quantum state. The condensate is described by the order parameter $\Delta e^{i\phi}$ where $\Delta^2$ gives the density of condensed pairs $\rho_s$ and φ is the phase. This macroscopic quantum state is characterized by exceptional manifestation of the quantum order: zero resistivity [1], perfect diamagnetism [2], quantization of magnetic flux [3,4] and quantum interference effects [5]. The wavefunction of the condensate decays exponentially as we go from the surface of the material to the vacuum with the Pippard coherence length $\xi_0$ [6] and the magnetic field decays exponentially [7] as we go from the surface into the material with the London penetration length $\lambda = \left( \sqrt{\frac{\rho_s}{m}} \sqrt{\frac{q}{\varepsilon_0 c^2}} \right)^{-1}$ .

The formation of the condensate made of electron pairs has been described by the BCS theory [8]. The key point of the BCS theory is that the formation of the condensate is due to the fact that electrons actually are not free particles but they are interacting; however the interaction is much smaller than the Fermi energy. In this weak coupling limit the interacting electrons are replaced by Landau quasiparticles. The very small electron-electron attraction triggers the formation of pairs of quasiparticles, with zero momentum and zero spin. The standard BCS theory assumes that the electron-phonon interaction provides the mechanism for the pairing however the pairing can also be mediated by electronic excitations (excitonic or plasmon mechanisms) in the low density limit.

In the weak coupling limit the critical temperature $T_C$ is related with the energy needed to break the pair $2\Delta_0$:

$$K_B T_c = \frac{2\Delta_0}{3.52} \qquad (1)$$

where $\Delta_0$ is the superconducting energy gap. The critical temperature (and the gap) is given by:

$$T_c = \frac{0.36}{f} \frac{T_F}{k_F \xi_0} \qquad (2)$$

where $T_F$ is Fermi temperature, $k_F = 2\pi/\lambda_F$ is the wavevector of electrons at the Fermi level, $\xi_0$ is the coherence length of the condensate that is related with the size of the pair and f is a measure of the deviation from the weak coupling limit $f = \frac{2\Delta_0}{3.52 K_B T_C}$,

The BCS approximations are valid for a 3D metal with critical temperature close to zero Kelvin. In the strong coupling limit the critical temperature for the many body superconducting phase remains low since the pairs form Bose particles at high temperature but the phase coherence of the Bose condensate occurs only at low temperature.

The discovery of high $T_c$ superconductivity [9] in copper oxide perovskites, with a record of $T_C \sim 150K$ in $HgBa_2Ca_2Cu_3O_{8+y}$ [10] has clearly shown that the superconducting condensate can be formed beyond the standard BCS approximations.

The mechanism driving the superconducting state from the range $0 < T_C < 23K$ of metals and alloys to the high temperature range

---

$20<T_c<150K$ of doped cuprate oxide perovskites, i. e., enhancing critical temperature by a factor ~10, is the object of this work.

The cuprate perovskites are heterogeneous materials formed by

1) metallic bcc $CuO_2$ layers intercalated between
2) insulating fcc $AO_{1-x}$ layers (A=Ba,Sr,La,Nd,Ca,Y.) [11-13] and
3) charge reservoir layers where the chemical dopants are stored.

Therefore the 3D superconducting order is realized in a superlattice of 2D metallic layers where the layers are separated by a spacing of the order of the Pippard coherence length. The pairs can jump between the layers while the single electrons are confined in the layers. In fact the pairs have a hopping of the order of $D_0 t_p^2$ between the 2D layers, where $D_o$ is the density of states at the Fermi level, $t_p$ is the single particle transverse hopping between the layers.

The critical temperature of the 3D condesate is controlled by the parameters of the 2D electron gas: the density $\rho=1/\pi (r_s a_B)^2$ where $r_s$ is the electron density parameter and $a_B$ is the Bohr radius; the Fermi wavevector $k_F=\sqrt{2}/(r_s a_B)$ and the Fermi energy $E_F(Ry) =2/(m^* r_s^2)$, and it is given by:

$$K_B T_c = 0.36 \frac{2}{m^* r_s} \frac{a_B}{f\sqrt{2}\xi_0} \text{ (Ry)} \qquad (3)$$

Uemura et al. and Keller et al. [14-16] have measured $\rho_s/m^*$ from the London penetration depth in different cuprate perovskites at a fixed doping, showing the linear relation for $T_c$ versus $\rho_s/m^*$.

Within the BCS approximations the critical temperature increases by increasing the electron-electron attraction, and by decreasing the size of the pairs, i. e., the coherence length $\xi_0$ of the superconducting phase.

However the BCS approximations breaks down in the strong coupling regime where the electron-electron attraction is larger than the Fermi energy. In this extreme limit all electrons form localized pairs (LP). These local pairs are formed below the high temperature $T_p$ however the superconducting critical temperature $T_c$ occurs at low temperature where the local pairs Bose condense and in the strong coupling the factor $f>>1$. Therefore the critical temperature $T_c$ reaches a maximum in a optimum intermediate coupling (OIC) regime where $k_F\xi_0\sim 2\pi$ [17].

## 2. THE METALLIC HETEROGENEOUS PHASE

The heterogeneous structure of a $Cu^{2+}$ cuprate perovskite is shown in Fig. 1. The $CuO_2$ layers form a fcc layer of a tetragonal structure with crystallographic axis $a_t =b_t=3.94$ Å. The Cu ion form a square piramid or bipyramid with planar Cu-O distance R=1.97 Å and axial Cu-O(A) distance 2.4-2.6 Å due to cooperative Jahn Teller effect for the $Cu^{2+}$ $3d^9$ ion that removes the degeneracy of the $Cu(3d_{x^2-y^2})$, $m\ell=2$, and $Cu(3d_{3z^2-r^2})$, $m\ell=0$, orbital.

The bcc $CuO_2$ layers are intercalated between insulating rocksalt fcc AO layers, This second material fit in the heterostructure by rotating its orthorhombic axis $a_0=b_0$ by $45^0$ and for the distance $R(AO)= R(Cu-O) \sqrt{2} = 5.57$ Å.

The electronic structure of the $CuO_2$ plane is a charge transfer insulator with a half filled valence band. The covalency of the Cu-O bond is very high and the single hole per Cu ion is both in the $Cu(3d_{x^2-y^2})$ orbital and/or in the molecular orbital combination of the oxygen 2p orbitals of local $b_1$ symmetry $L(b_1)= \frac{1}{2}\left(p_{x_1} - p_{y_2} - p_{x_3} + p_{y_4}\right)$. There is a strong local Coulomb repulsion for two holes (d) in the same Cu 3d orbital, $U_{dd}\sim 6$ eV, that gives a Mott-Hubbard gap $\Delta_H=U_{dd}$ for the charge transfer $(\underline{d}_x+\underline{d}_x)\to(0+\underline{d}_x\underline{d}_x)$ where $\underline{d}_x$ indicates a hole in the $Cu(3d_{x^2-y^2})$ orbital.

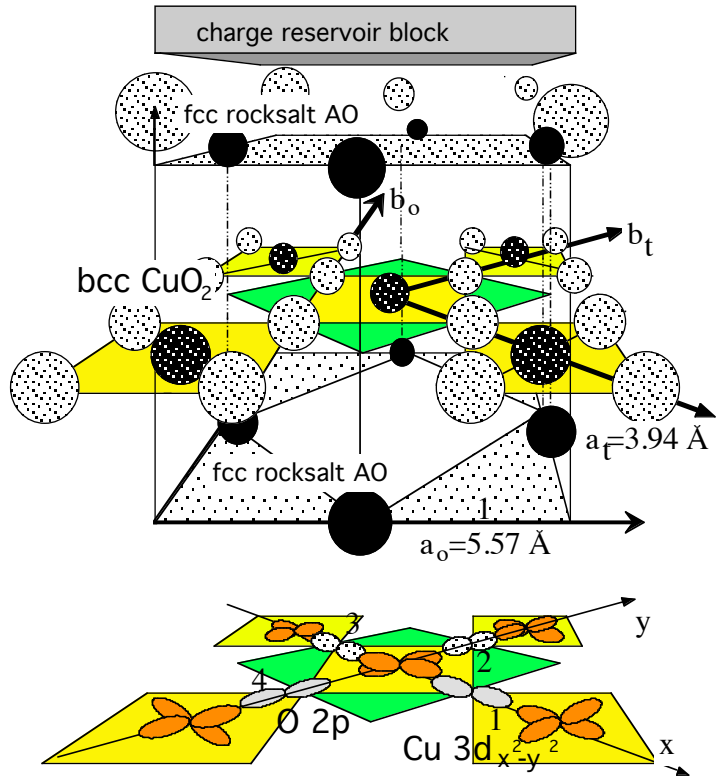

Fig. 1 The heterostructure of a $Cu^{2+}$ cuprate perovskite and the molecular orbitals forming the electronic structure of the $CuO_2$ plane.

The gap for the electron transfer of a hole from the $Cu(3d^9)$, or $\underline{d}_x$ to the oxygen orbital $\underline{L}(b_1)$, $(\underline{d}+\underline{L})\to(0+\underline{dL})$, is smaller than the Hubbard gap. This charge transfer gap is given by $\Delta_{CT} = \varepsilon_L - \varepsilon_d + U_{dL}$, where in the final state configuration there is a Coulomb repulsion $U_{dL}$ between a hole on Cu and a hole on the nearest oxygen, and $\varepsilon_L - \varepsilon_d$ is the energy separation between O(2p) and Cu($3d_{x^2-y^2}$) orbital. The relevant local inter-atomic Coulomb repulsion $U_{dL} \sim 2$ eV has been determined by joint x-ray photoemission and x-ray absorption and the optical gap for the insulating compound $\Delta_{CT} \sim U_{dL}$ since $\varepsilon_L - \varepsilon_d \sim 0$ [18].

The metallic phase in the CuO$_2$ plane is obtained by two separate steps in the design of the material: first, chemical dopants that play the role of acceptors and pump electrons from the CuO$_2$ plane, are introduced in the charge reservoir blocks; second, multiple substitutions of metallic ions A (A=Ba,Sr,La,Nd,Ca,Y.,) in the rocksalt layers are made in such a way to change the average ionic radius <$r_A$> of the rocksalt layers. In cuprates with multiple CuO$_2$ layers the rocksalt layers between the copper planes loose completely their oxygen ions. Doping introduces holes in the O 2p orbital (L) and a single hole remains in the Cu site [19]. However the symmetry of the molecular orbital for the added hole have a mixed symmetry with a component with local a$_1$ symmetry L(a$_1$)= $\frac{1}{2}(p_{x_1} + p_{y_2} - p_{x_3} - p_{y_4})$ mixed with d$_z$ = Cu($3d_{3z^2-r^2}$) m$\ell$=0 [20-22]. These results have shown that the symmetry of the doped holes is not the pure m$\ell$=2 symmetry of the antiferromagnetic insulator at half filling. Therefore the doped holes in the metallic phase are associated with a local lattice distortions (LLD) mixing states with different orbital momentum. These LLD distortions are expected for a pseudo Jahn Teller electron lattice interaction of the doped holes [23]. Here the key point has been to show that the electronic correlation lower the Jahn-Teller energy separation $\Delta_{JT}$ between the states with m$\ell$=2 and m$\ell$=0 symmetry from about 1.5 eV to about 0.5 eV since the Coulomb repulsion $U_{dL}$ for the configuration L(a$_1$)d$_z$ is much smaller than for L(b$_1$)d$_x$.

The 2D electron gas in the CuO$_2$ plane of cuprate perovskites is therefore a strongly correlated electron gas described by the Hubbard Hamiltonian. Moreover there is a relevant electron lattice interaction of the type of cooperative pseudo Jahn-Teller coupling of charges with Q$_2$-type local modes. This can be described by the Holstein Hamiltonian with a next-near neighbour hopping integral t'. Therefore its metallic phase is described by the Hamiltonian:

$$H = H_{el} + H_U + H_{ph} + H_I =$$

$$= -t \sum_{<i,j>\sigma}\left(c^+_{i\sigma}c_{j\sigma}\right) - t'\sum_{<<i,j>>\sigma}\left(c^+_{i\sigma}c_{j\sigma}\right) + U\sum_i n_{i\uparrow}n_{i\downarrow}$$

$$+ \omega_0 \sum_q a^+_q a_q + g\omega_0 \sum_{i,q} c^+_i c_j \left[a^+_q a_q\right] - \mu_0 \sum_{i\sigma} n_{i\sigma} \qquad (4)$$

The first two terms describe the itinerant charges in a 2D square lattice simulating the $CuO_2$ plane where t is the electron transfer integral between nearest-neighbor sites $<i,j>$ and t` is the electron transfer integral between next-nearest-neighbor sites $<<i,j>>$, $n_{i\sigma}=c_{i\sigma}^+c_{i\sigma}$ is the local electron density, $c_{i\sigma^*}$ denotes the electron creator operator at site i.

The third term is the Hubbard Hamiltonian describing the electronic correlation in the $CuO_2$ plane. The Hubbard term induces a mass renormalization of a factor of the order of 5 giving an effective mass in the $(\pi,\pi)$ direction, $m^*/m_0 \sim 2$.

The coupling of the charges with local lattice distortions (LLD) of the $CuO_4$ unit can be described by the Holstein Hamiltonian ($H_{ph}+H_I$). The position of the lattice site is indicated by $R_i$ and $a_q^+$ represents the creation operator for phonon with wavevector q, $\omega_0$ is the frequency of the optical local phonon mode and g indicates the coupling of the charge with this local lattice mode. This term describes the weak electron-phonon interaction while for g sufficiently large the charge is coupled with local lattice distortions.

The local lattice distortion Q follows the equation

$$\ddot{Q}_q = -\omega_0^2 Q_q - g\omega_0\left(\frac{2\omega_0}{M\hbar}\right)^{1/2},$$

therefore the electron-lattice interaction ($g \neq 0$) provides a force $F = -g\omega_0\left(\frac{2\omega_0}{M\hbar}\right)^{1/2}$ that induces a displacement $\Delta Q = \frac{-g}{\omega_0}\left(\frac{2\omega_0}{M\hbar}\right)^{1/2}$ of the equilibrium position. The local lattice distortion (LLD) appears when $\Delta Q$ becomes larger than zero energy vibration amplitude. In the present square lattice it is possible to identify the electron lattice coupling constants $\lambda_1 = \frac{g^2\omega_0}{2t\,d}$, $\lambda_2 = \frac{g^2\omega_0}{2t'd}$ where d=2 for a 2D electron gas. We have found that in the cuprate perovskites, while the first coupling constant is in the weak coupling limit, $\lambda_1<1$, the second one is in the intermediate-strong coupling limit $1<\lambda_2<2.5$ and it is expected to give local lattice distortions. In this situation we are in an intermediate regime where charges trapped into LLD coexist with itinerant charges. This situation is expected to occur in special cases in the intermediate electron-lattice coupling regime.

The experimental evidence for LLD due to pseudo JT electron-lattice interaction (JT-LLD) was provided by the presence of two different types of doped holes in the oxygen orbital [20-23]:

1) $\underline{L}(a_1)$ of partial $a_1$ symmetry, mixed with Cu $3d_{3z^2-r^2}$ (orbital angular momentum $m\ell=0$) and
2) $\underline{L}(b_1)$ of $b_1$ symmetry mixed with Cu $3d_{x^2-y^2}$ (orbital angular momentum $m\ell=2$).

The pseudo JT-LLD should be associated with the doped holes $\underline{L}(a_1)$ since the $Q_2$ type lattice distortion forms molecular orbital of mixed $m\ell=2$ and $m\ell=0$ angular momentum. The search for these local lattice distortions motivated the Rome group to solve the incommensurate structural modulation of the $CuO_2$ plane in $Bi_2Sr_2CaCu_2O_{8.2}$ (Bi2212) by joint single crystal x-ray diffraction and EXAFS. We have found in 1992 that the pseudo JT-LLD get self organized in linear arrays, i.e., stripes [24-26]. The co-existing itinerant particles form rivers of charges and at the Erice workshop in 1992 [25] the scenario of superconducting stripes, where *"the free charges move mainly in one direction, like the water running in the grooves of a corrugated iron foil"*, was introduced for the first time in the field of high $T_c$ superconductors.

A heterogeneous phase of the matter is a generic phenomenon following doping of a high correlated antiferromagnetic insulating electronic system. The formation of a microscopic electronic phase separation with the formation of metallic droplets in an antiferromagnetic background was first shown in doped magnetic semiconductors [27]. Experimental evidence that at very low doping in the cuprates the doped holes segregate into strings of charges that play the role of domain walls between anti ferromagnetic domains, forming a glassy phase of strings, has been reported [28]. At higher doping if the counterions are mobile the system is unstable toward a macroscopic phase separation between macroscopic metallic domains and insulating antiferromagnetic domains [29,30].

The high $T_c$ superconductors are a special case of heterogeneous doped magnetic superconductors since in the metallic droplets we have the coexistence of doped charges in the weak coupling limit, phase A, with doped charges associated with local lattice distortions associated with pseudo Jahn Teller electron lattice interaction phase B. There is now clear experimental evidence that there are two types of doped charges in the cuprates [31]. The ordering of charges trapped by the pseudo Jahn-Teller LLD with an associated modulation of the orbital angular momentum gives stripes and orbital density waves.

The phase diagram of the metallic phase of high $T_c$ superconductors is usually given as a function of hole doping, measuring the distance from the antiferromagnetic (AF) insulating Mott Hubbard phase. In the two components 2D electron fluid it is necessary to measure the actual density of the itinerant component by using the electron density parameter $r_s$ measured by the Hall effect at low temperature.

The phase diagram of $La_{2-x}Sr_xCuO_4$ as a function of electron density parameter $r_s$ is shown in Fig. 2 (lower panel).

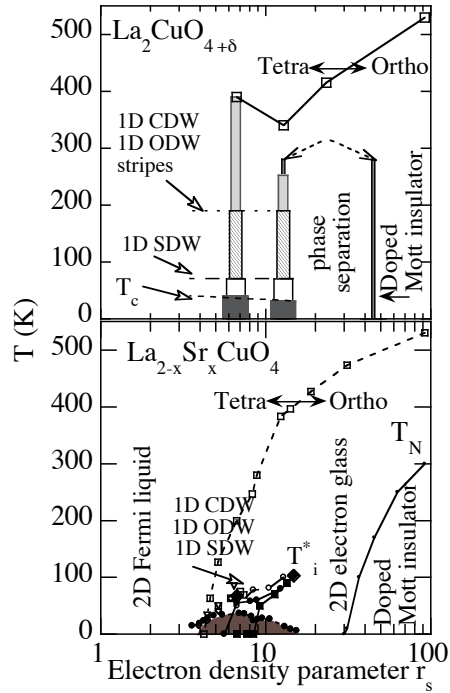

Fig. 2. Phase diagram of the $La_{2-x}Sr_xCuO_4$ (lower) and $La_2CuO_{4+\delta}$ (upper) as a function of electron density parameter $r_s$ of the itinerant 2D electron gas measured by Hall effect. Usual notations are used to denote different regimes of the phase diagram. The structural phase transition boundary between the orthorhombic and tetragonal phase is shown.

The system $La_{2-x}Sr_xCuO_4$ shows a complex phase diagram typical of a glassy system due to the random distribution of countercharges (Sr ions) in the block layers. The AF phase appears in the range $r_s>37$, and a spin glass phase appears for $37>r_s>15$. The metallic phase, $15>r_s>5$, shows a typical glassy phase with several crossover temperatures $T^*_i$ that depend on the measuring time of each experimental probe.

To understand the basic physics of the metallic phase of cuprate superconductors we need to study a simple crystalline system. This is provided by $La_2CuO_{4+\delta}$ (and $Bi_2Sr_2CaCu_2O_{8+\delta}$) where the itinerant holes in the $CuO_2$ plane are compensated by the negative charges carried by the mobile interstitial oxygen $\delta$ in the LaO layers (and in the BiO charge reservoir layer) that can get ordered. There is no frustrated phase separation regime in $La_2CuO_{4+\delta}$ due to mobile counterions, therefore it does not show the spin glass phase of the doped Mott insulator in the range $37>r_s>15$, where it shows the expected phase separation below about 300K between an insulating doped AF lattice ($r_s \sim 37$) and a metallic phase ($r_{s1} \sim 12$).

This first superconducting phase with $T_c=32$ K shows the universal 1D dynamical spin fluctuations below $T_{sdw} \sim 60$K and the 1D stripes [32], CDW and/or ODW, indicated by the ordering of local lattice distortions of the $CuO_6$ octahedra (tilting) below a critical temperature $T_{c0}=190$ K [33], as shown in Fig. 2 (upper panel).

A second stable phase with the highest critical temperature appears at $5<r_{s2}<6$. In this phase, oxygen ordering and charge and orbital stripes in the $CuO_2$ plane have been found by several experimental techniques probing different physical parameters: NMR revealing two different Cu sites [34], EXAFS solving the local $CuO_4$ rhombic distortions, characteristic of the pseudo Jahn Teller polarons below 150 K [35], x-ray [36] and electron diffraction [30]. The universal 1D dynamical spin fluctuations at $(\pi,\pi\pm\delta)$ and $(\pi\pm\delta,\pi)$, $\delta=0.105$, have been also observed in this regime below 60 K by inelastic magnetic neutron scattering (37).

In a recent work we have identified the short range incommensurate charge ordering in the $CuO_2$ plane reflected by a pattern of diffuse x-ray scattering peaks with wavevector $q_{CDW}=(0.208\mathbf{b}^*,0.29\mathbf{c}^*)$ and a coherence length of about 350 Å. The anharmonicity of this modulation is evident from large intensity of higher harmonics. By cooling the sample the diffuse charge ordering peaks show a temperature dependence reported in Fig. 3, where the square root of the intensity of the second harmonic peak is plotted.

In the second harmonic peaks we can well separate the charge ordering from the 3D oxygen ordering peaks. The square root of the intensity plotted in Fig. 3 gives the direct measure of the density of charge $\Delta\rho(q)$ ordered in the $CuO_2$ plane with wavevector q, that is the order parameter for the charge ordered phase. The solid line is a fit to the experimental intensities with an expression $\Delta\rho \propto (T-T_{co})^\alpha$ which represents a typical second order phase transition with $T_{co} \sim 188$ K. This effect is clearly due to formation of charge stripes in the $CuO_2$ plane since the oxygen mobility is frozen below 200K. In fact the 3D oxygen ordering has already been established at higher temperature (in the range 270-230 K) in the system as evidenced by temperature evolution of the resolution-limited diffraction peaks.

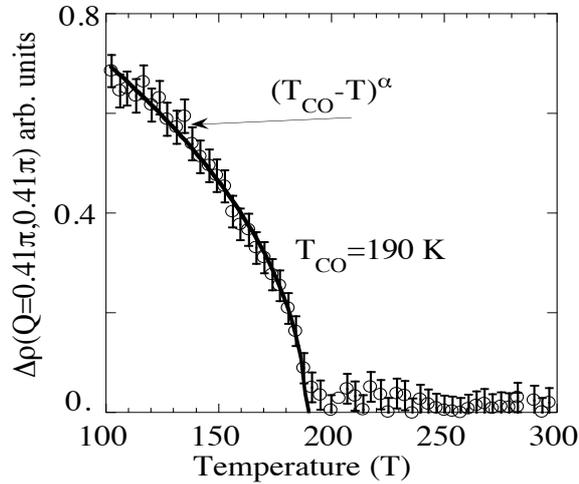

Fig. 3. Temperature dependence of the order parameter for the stripe formation with wavevector q(0.4π,0.4π) in $La_2CuO_{4.1}$. The fit shows a critical temperature $T_{co}$=190 K for charge ordering.

The second harmonic of the charge modulation at 0.416**b**\* has the same wavevector of the nesting vector at $2k_F \sim 0.4\mathbf{b}^*$ or ~(0.4π,0.4π) observed by Saini et al. [18] in the Fermi surface of Bi2212 that induces the suppression of the spectral weight at selected spots in the k space and gives a broken Fermi surface. The 3D ordering of dopants stabilizes the orthorhombic phase, as shown in Fig. 1, and the symmetry of the $CuO_2$ plane is broken. The ordering of stripes in the **b** direction gives a superconducting phase in a broken symmetry, with a broken Fermi surface. The Fermi surface is therefore formed not by closed circles but by segments and the "mini-gaps" in the density of states due to the 1D superlattice of stripes [19-20] give origin to the "pseudo-gap" scenario. In this stripe scenario the amplification of the superconducting critical temperature is realized by tuning the Fermi level at a "shape resonance" of the superlattice [19-20]. The pairing mechanism is mediated by charge fluctuations in a metal with the anomalous dielectric constant typical of an anomalous Fermi fluid at $r_s$>4 [17].

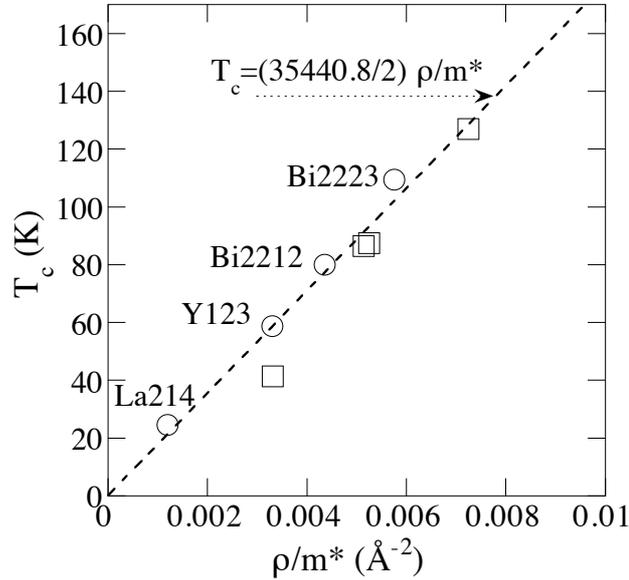

Fig. 4. The critical temperature of different cuprate perovskites at fixed doping at $\delta_1 \sim 0.1$ (open circles) and $\delta_2 \sim 0.16$ (open squares) as function of the ratio $\rho/m^*$, where $\rho$ is the condensate density and $m^*$ the effective mass. The dashed line shows the calculated $T_c$ using formula (5) for the coherence length $\xi_0$ of the order of the distance between particles $<d>=2r_s a_B$.

In this regime a generalized BCS scheme remains valid and the dynamical pairing described by the BCS is still possible up to the point where the size of the pairs of quasi-particles is of the order of the wavelength of the electrons at the Fermi level $\xi_0 \sim \lambda_F$. In this regime the coexistence of local pairs (bosons) and fermions in the normal phase is expected. The proximity to Wigner localization gives a metal with a negative dielectric permitivity $\varepsilon(\omega,q) \leq 0$, $q \neq 0$, $\omega \rightarrow 0$, predicted by Dolgov and Ginsburg, needed for high $T_c$ with $k_F \xi_0 = 2\pi$. The highest critical temperature is reached for $\xi_o$ close to the average distance between two electrons $<d>=2r_s a_B$ in a 2D electron gas. This is the highest critical temperature possible in an extended BCS scheme in the intermediate coupling:

$$T_c (K) = (35440.8/f) \rho/m^* \qquad (5)$$

where $\rho$ is measured in $Å^{-2}$, and $m^*$ is the effective mass. In this limit the critical temperature depends only on the ratio $\rho/m^*$ and using the phenomenological value, f=2, from tunneling data at optimum doping, the critical temperature has been calculated as a function of $\rho_s/m^*$ in ref. 17. The predicted temperature is plotted in Fig. 4 and it is in very good agreement with the experimental data.

The expected density of charge carriers is constant but the condensate density is different in different cuprate perovskites. Therefore there is a term in the Hamiltonian of the metallic phase of the $CuO_2$ layers, that changes the electronic structure and drives the critical temperature and the condensate density. The object of this work is the identification of the term in the Hamiltonian that drives the $CuO_2$ plane to the optimum intermediate coupling regime giving high $T_c$ superconductivity.

## 3. THE $T_C(\delta)$ PHASE DIAGRAM OF BI2212

After the discovery of HTcS the physics of cuprates was described by a generic phase diagram $T_c(\delta)$ where the critical temperature is plotted as a function of doping $\delta$, i. e., a measure of the charge density and the distance from the Mott Hubbard insulator at $\delta=0$. By increasing $\delta$ the system goes through quite different states. At low doping the doped holes form a disordered electron glass. At very high doping a normal metallic phase appears. The high $T_c$ superconducting phase appears between these two phases. In 1993 we presented the phase diagram $T_c(\delta)$ for Bi2212, shown in Fig. 5. The doping was measured in units of $\delta_0$, the critical density for the insulating commensurate JT charge ordered crystal at $\delta_0=1/8$ in $La_{1-x}Ba_xCuO_4$ This electronic crystal is in competition with superconductivity and it is at the origin of the huge suppression of the superconducting critical temperature in $La_{1-x}Ba_xCuO_4$ at $x=1/8$. The long range Coulomb interaction between charge trapped in pseudo JT-LLD is expected to play a key role in the formation of the ordered phase of localized charges. Therefore we call this phase a generalized Wigner commensurate JT polaron crystal (CPC).

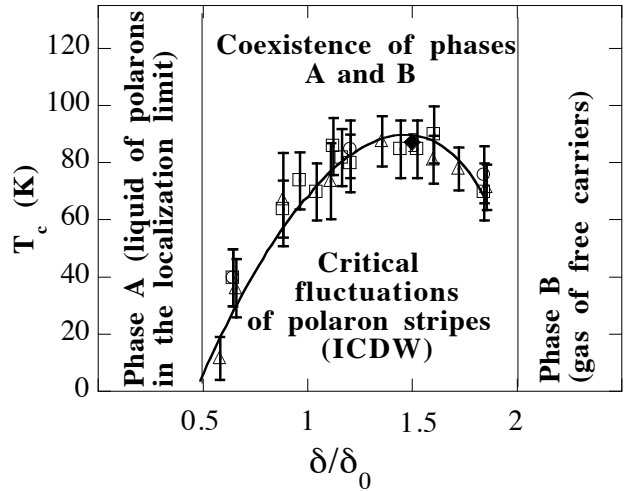

Fig. 5. The phase diagram $T_c(\delta)$ for Bi2212. The charge density is measured in units of the critical doping $\delta_0=1/8$ for the expected commensurate polaron crystal (CPC) also if in this family by changing the doping we do not cross the CPC phase. We observe 1) an insulating phase A of an electron gas in the localization limit ($k_Fl<1$) for $\delta/\delta_0<0.5$; 2) a homogeneous metallic phase B at $\delta/\delta_0>2$ and 3) a region of co-existence of the two different phases, $0.5<\delta/\delta_0<2$: a 1D-polaronic incommensurate charge density waves and the superconducting phase.

The CPC was not observed at $\delta=\delta_0$ in the phase diagram $T_c(\delta)$ of Bi2212 shown in Fig. 5 where only a weak minimum of $T_c$ appears at $\delta_0=1/8$. By decreasing the temperature, below about $T^*=T_{co}=130$ K, the system form an inhomogeneous phase where a one-dimensional (1D) incommensurate polaronic charge density wave (ICDW) or polaronic stripes coexist with free carriers. In the underdoped phase, $0.5<\delta/\delta_0<1$, the charge density of free carriers is smaller than that of JT polarons in the ICDW. In the high doping phase, $1<\delta/\delta_0<2$, the charge density of free carriers is larger than that of JT polarons in the ICDW. This particular ICDW does not suppress but promotes the pairing of the free carriers below $T_c$. In fact in this unexpected metallic phase of condensed matter, a superlattice of quantum mesoscopic stripes of width L, the chemical potential is tuned to a "shape resonance". The "shape resonance" occurs when the de Broglie wavelength of electrons at the Fermi level $\lambda_F$ ~nL/2. The measure of the stripe width L in 1993 has established the presence of the "shape resonance" L~$\lambda_F$ at optimum doping. At the shape resonance the chemical potential is tuned to the bottom of a superlattice subband and therefore to a narrow peak in the density of states (DOS). At optimum doping a BCS-like superconductivity is observed ($2\Delta/T_c \sim 5$) and the highest $T_c$ is reached where the chemical potential is tuned to this narrow DOS peak via the calculated "shape resonance" effect on the superconducting gap. A patent has been granted with priority date 7 Dec 1993 for a method of $T_c$ amplification via the "shape resonance" effect in new materials formed by a superlattice of quantum wires. A very good agreement with experimental data has been found for the calculated critical temperature plotted as a solid line in Fig. 5 assuming the pairing mechanism mediated by charge fluctuations in a superlattice of quantum wires at the "shape resonance" [17].

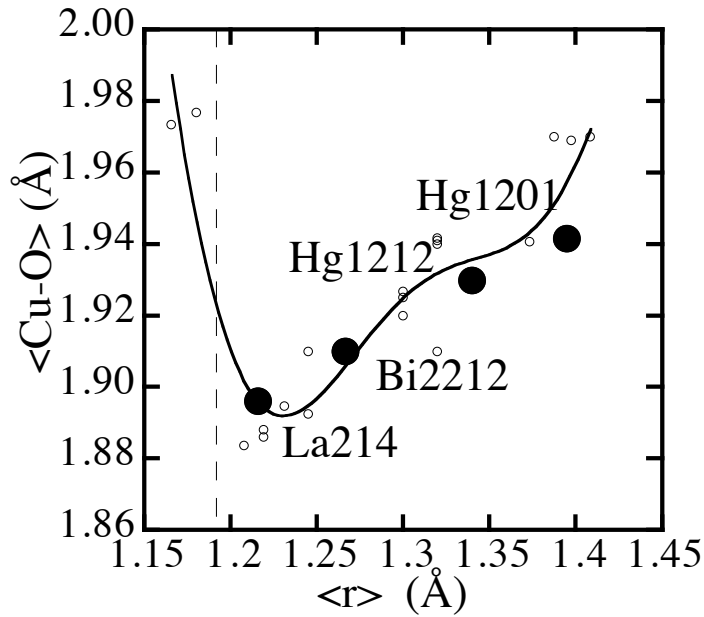

Fig. 6. The average <Cu-O> bond length measured by Cu K-edge EXAFS as a function of the average radius <r> of metal ions in the rocksalt layers.

## 4. THE MICRO STRAIN QUANTUM CRITICAL POINT

The electron-lattice interaction g of the pseudo JT type in the cuprates is controlled by the static distortions of the $CuO_4$ square plane and the Cu-O(apical) distance. In fact $g=\Psi(Q)\phi(\Delta_{JT})\gamma(\beta)$ where Q is the conformational parameter for the distortions of the $CuO_4$ square, like the rhombic distortion of $CuO_4$ square; $\beta$ is the dimpling angle given by the displacement of the Cu ion from the plane of oxygen ions; and $\Delta_{JT}$ is the JT energy splitting that is controlled by the Cu-O(apical) bond length.

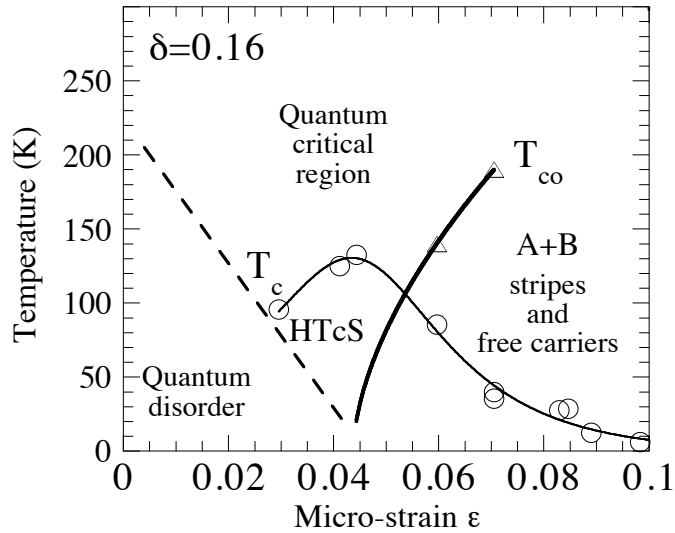

Fig. 7. The critical temperature for charge ordering $T_{co}$ and the superconducting critical temperature $T_c$ as function of the micro strain $\varepsilon$ at optimum doping $\delta=0.16$.

There is an external field acting on the $CuO_2$ plane of the cuprates that controls $g(\eta)$ via the micro-strain of the $CuO_2$ lattice due to the compressive stress generated by the lattice mismatch between the metallic bcc $CuO_2$ layers and the insulating rocksalt fcc AO layers [40,41]. The bond-length mismatch across a block-layer interface is given by the Goldschmidt tolerance factor $t=[r(A-O)]/\sqrt{2}[r(Cu-O)]$ where $[r(A-O)]$ and $[r(Cu-O)]=d_0$ are the respective equilibrium bond lengths in homogeneous isolated parent materials A-O and $CuO_2$ [42]. The hole doped cuprate perovskite heterostructure is stable in the range $0<t<0.9$ that corresponds to a mismatch $\eta=1-t$ of $0<\eta<10\%$.

We have focused our attention to 3 main cuprate perovskite systems 1) Hg1212, 2) Bi2212 and 3) $La_2CuO_{4+y}$. In these materials the hole concentration in the $CuO_2$ plane is controlled by the oxygen doping in the charge reservoir blocks. The stress due to the mismatch or the chemical pressure acting on the $CuO_2$ plane is controlled by the average ionic radius in the rocksalt layers. The stress increases going from Ba to Sr to La. We have measured the average Cu-O bond lengths by Cu K-edge EXAFS, a local structural probe, and shown in Fig. 6 as a function of average ionic radius of metallic ions in the rocksalt layers <r>. Decreasing <r> is equivalent to a anisotropic chemical pressure acting on the $CuO_2$ plane. We define the local or micro strain of the CuO in plane bond length $\varepsilon=2(1-\langle Cu\text{-}O\rangle/d_0)$, where $d_0=1.97$ Å is the equilibrium Cu-O distance at doping $\delta=0.16$ in many different systems.

In the cuprate perovskites the micro-strain $\varepsilon$ drives the system to a quantum critical point $g_c(\varepsilon_c)$ for the formation of a superlattice of quantum stripes. The stripes of local lattice distortions are detected by x-ray diffraction above a critical micro-strain $\varepsilon_c \sim 4\%$.

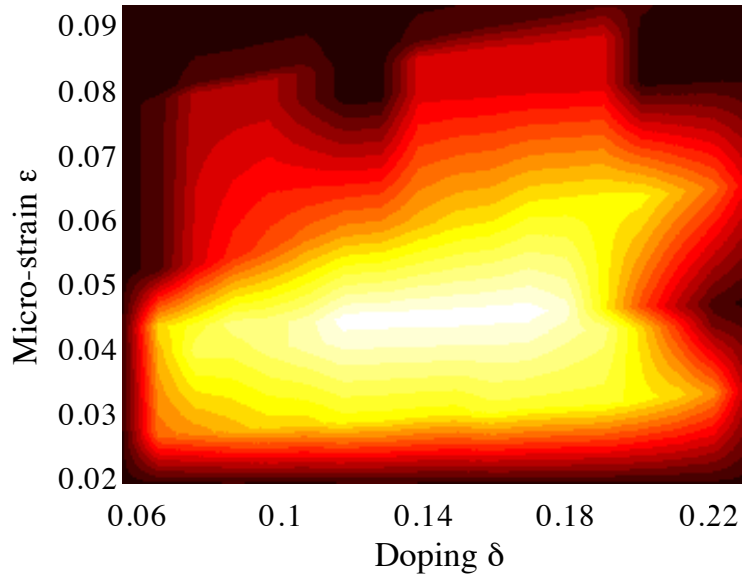

Fig. 8. The superconducting critical temperature $T_c$ plotted in a half-tone scale scale (from $T_c=0K$, black, to $T_c \sim 135K$, white) as a function of the micro-strain $\varepsilon$ and doping $\delta$. The maximum $T_c$ occurs at the critical point $\delta_c=0.16$, and $\varepsilon_c=0.04$.

We have plotted in Fig. 7 the variation of the critical temperature for charge ordering $T_{co}$ and the superconducting critical temperature $T_c$ as function of the micro-strain at optimum doping $\delta=0.16$.

In Fig. 8 we report the critical temperature $T_c$ in a color plot (the critical temperature increases from black, $T_c=0K$, to white, the maximum $T_c \sim 135K$) as a function of the micro-strain $\varepsilon$ and doping $\delta$ for all superconducting cuprate families. The figure shows that the maximum $T_c$ occurs at the critical point $P(\delta_c=0.16;\varepsilon_c\sim 0.04)$.

From these data we can derive a qualitative phase diagram for the normal metallic phase of all cuprate perovskites that give high $T_c$ superconductivity that is shown in Fig. 9.

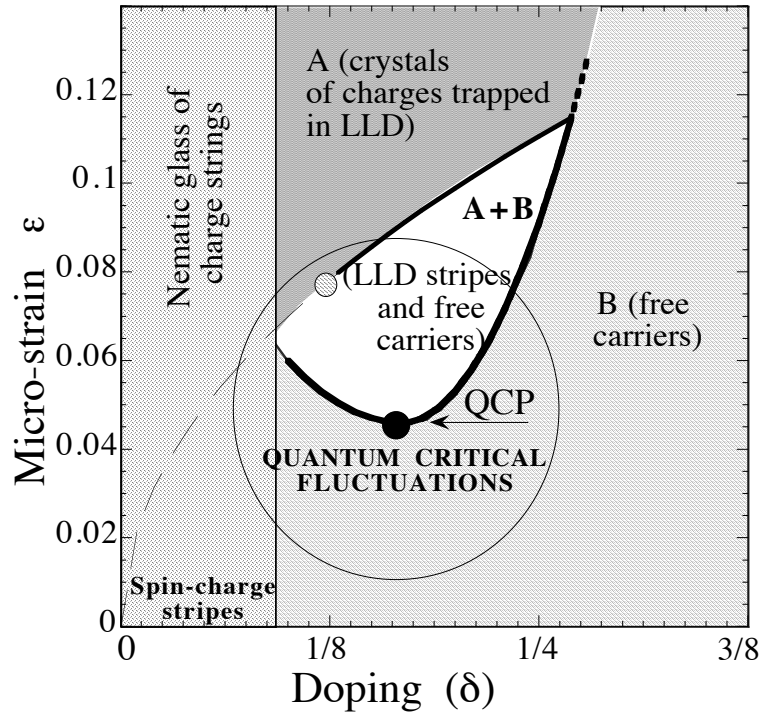

Fig. 9. The phase diagram of the normal phase of doped cuprate perovskites as a function of micro-strain on the Cu-O(planar) bond and doping. The high $T_c$ superconductivity occurs in the region of quantum fluctuations around the micro-strain quantum critical point QCP.

This phase diagram solves the long standing puzzle of the phase diagram of the normal phase of the cuprates. There was a hidden physical parameter, the micro-strain, that triggers the electron-lattice interaction at a critical value for the onset of charges trapped into pseudo JT-LLD. The doping of the strained antiferromagnetic lattice forms both free carriers and charges trapped into the JT-LLD above the critical micro-strain $\varepsilon_c$. For $\varepsilon > \varepsilon_c$, as it was discussed for the case of oxygen doped Bi2212 and La124, the systems show a quasi first order phase transition as a function of doping.

The quantum critical point QCP is well defined at constant finite doping as a function of the micro-strain as it is shown in Fig. 8. Direct experimental evidence for quantum critical local lattice fluctuations has been obtained by measuring the dynamical fluctuations of the Cu-O bond at a high temperature $T_H > T_{co}$ in all families of cuprates ($T_H \sim 200K$).

In conclusion we have deduced a phase diagram for the superconducting phases where $T_c$ depends from both doping and micro-strain. The anomalous normal phase of cuprate superconductors is determined by an inhomogeneous phases with co-existing polaronic stripes and itinerant carriers that appears for an electron lattice interaction larger than a critical value. Fluctuations of lattice-charge stripes appear in this critical regime. The micro-strain drives the electron lattice interaction to a QCP of a quantum phase transition [43]. Near this QCP the stripes get self organized in a superlattice of quantum wires of charges trapped into JT-LLD that co-exist with free carriers. This superlattice forms an array of superstripes where the chemical potential is tuned to a shape resonance. The plot $T_c(\varepsilon)$ reaches the highest temperature at the critical point $\varepsilon_c$.

This research has been supported by INFM, by MURST - Programmi di Ricerca Scientifica di Rilevante Interesse Nazionale, and by "Progetto 5% Superconduttività del CNR.

**References**
[1]    H. Kamerling Onnes Comm. *Phys. Lab. Univ. Leiden* Nos. **119, 120, 122** (1911).
[2]    W. Meissner & R. Ochsenfeld *Naturwiss* **21**, 787 (1933).
[3]    F. London *Superfluids* J. Wiley and Sons Inc. New York, 1950 Vol. **1** pag. 152
[4]    R. Doll & M. Nabauer *Phys. Rev. Lett.* **7**, 51 (1961); B. S. Deaver Jr. and W. M. Fairbank *Phys. Rev. Lett.* **7**, 43 (1961).
[5]    B. D. Josephson *Physics Lett.* **1**, 251 (1962).
[6]    P. B. Allen & B. Mitrovic *Solid State Physics* **37**, 1 (1982).
[7]    H. London, & F. London *Proc. Roy. Soc. (London)* **A149**, 71 (1935); *Physica* **2**, 341 (1935).
[8]    J. Bardeen, L. N. Cooper, and J. R. Schrieffer *Phys. Rev.* **108**, 1175 (1957).
[9]    J. G. Bednorz, & K. A. Müller *Z. Phys.* B **64,** 189(1986); J. G. Bednorz and K. A. Müller *Rev. Mod. Phys.* **60** 565 (1988).
[10]   C. W. Chu et al. *Nature* **365**, 323 (1993).
[11]   J. -M. Triscone, et al *Physical Review Letters* **63**, 1016 (1989).
[12]   Qi Li, T. Venkatesan & X. X. Xi *Physica C*, **190**, 22 (1991).
[13]   M. Holcomb, J. P. Collman, & W. A. Little *Physica C*, **185-189**, 1747(1991)
[14]   Y. J. Uemura et al. *Phys. Rev. Letters* **62**, 2317 (1989); *Phys. Rev. B* **38**, 909 (1988); *Phys. Rev. Lett.* **66**, 2665 (1991);*Nature* **364**, 605 (1993).
[15]   H. Keller, et al. *Physica* (Amsterdam) **185-189C**, 1089 (1991); T. Schneider & H. Keller *Phys. Rev. Lett.* **69**, 3374 (1992);
[16]   Ch. Niedermayer et al. *Phys. Rev. Lett.* **71**, 1764 (1993).
[17]   A. Bianconi *Sol. State Commun.* **91**, 1 (1994).
[18]   A. Bianconi, A. Clozza, A. Congiu Castellano, S. Della Longa, M. De Santis, A. Di Cicco, K. Garg, P. Delogu, A. Gargano, R. Giorgi, P. Lagarde, A. M. Flank, and A. Marcelli in *International Journal of Modern Physics B* **1**, 853 (1987).
[19]   A. Bianconi, A. Congiu Castellano, M. De Santis, P. Rudolf, P. Lagarde, A. M. Flank, and A. Marcelli, *Solid State Communications* **63**, 1009 (1987).
[20]   M. Pompa, S. Turtù, A. Bianconi, F. Campanella, A. M. Flank, P. Lagarde, C. Li, I. Pettiti, and D. Udron, *Physica* **C185-189**, 1061-1062 (1991)**.**
[21]    M. Pompa, P. Castrucci, C. Li, D. Udron, A. M. Flank, P. Lagarde, H. Katayama-Yoshida, S. Della Longa, and A. Bianconi, *Physica* **C184,** 102-112 (1991).
[22]   A. Bianconi M. Pompa, S. Turtu, S. Pagliuca, P. Lagarde, A. M. Flank, and C. Li *Jpn. J. Appl. Phys.* **32** (Suppl. 32-2) 581 (1993).
[23]   Y. Seino, A. Kotani, and A. Bianconi *J. Phys. Soc. Japan* **59**, 815 (1990).
[24]    A. Bianconi, S. Della Longa, M. Missori, I. Pettiti, and M. Pompa in *Lattice Effects in High-T$_C$ Superconductors, edited by* Y. Bar-Yam, T. Egami, J. Mustre de Leon and A. R. Bishop; World Scientific Publ., Singapore, (1992) pag. 6.

[25]   A. Bianconi, in "*Phase Separation in Cuprate Superconductors*" ed. by K. A. Muller & G. Benedek, World Scientific Singapore (1993) p. 352; ibidem pag. 125.
[26]   A. Bianconi, S. Della Longa, M. Missori, I. Pettiti, and M. Pompa and A. Soldatov *Jpn. J. Appl. Phys.* **32** suppl. 32-2, 578 (1993).
[27]   E. L. Nagaev Sov. Jour. JEPT Lett. 16, 558 (1972); V. A. Kaschin and E. L. Nagaev *Zh. Exp. Teor. Phys.* **66**. 2105 (1974).
[28]   J. H. Cho, F. C. Chu and D. C. Johnston, *Phys. Rev. Lett.* **70**, 222 (1993).
[29]   P. C. Hammel, A. P. Reyes, Z. Fisk, M. Takigawa, J. D. Thompson, R. H. Heffner, S. Cheong, and J. E. Schirber *Phys. Rev. B* **42**, 6781 (1990).
[30]   J. C. Grenier, N. Lagueyte, A. Wattiaux, J. -P. Doumerc, P. Dordor, J. Etourneau and M. Pouchard, *Physica C* **202**, 209. (1992).
[31]   D. Mihailovich, and K. A. Müller, *High $T_c$ Superconductivity: Ten years after the Discovery* (Nato ASI Series-Applied Sciences, *Vol*. 343) ed E. Kaldis, E. Liarokapis, K. A. Müller, (Dordrecht, Kluwer) pag. 243-256.
[32]   B. O. Wells, Y. S. Lee, M. A. Kastner, R. J. Christianson, R. J. Birgeneau, K. Yamada, Y. Endoh, and G. Shirane *Science* **277**, 1067-1071 (1997).
[33]   X. Xiong, P. Wochner, S. C. Moss, Y. Cao, K. Koga and N. Fujita, *Phys. Rev. Letters* **76**, 2997 (1996).
[34]    B. W. Statt, P. C. Hammel, Z. Fisk, S. W. Cheong, F. C. Chou, D. C. Johnston and J. E. Schirber, *Phys. Rev. B* **52**, 15575 (1995).
[35]   P. G. Radaelli, J. D. Jorgensen, R. Kleb, B. A. Hunter, F. C. Chou and D. C. Johnston*, Phys. Rev. B* **49**, 6239 (1994).
[36]   A. Lanzara, N. L. Saini, A. Bianconi, J. L. Haemann, Y. Soldo, F. C. Chou, and DC. Johnston *Phys. Rev. B* 55, 9120 (1997).
[37]   Y. S. Lee, R. J. Birgeneau, M. A. Kastner, Y. Endoh, S. Wakimoto, K. Yamada, R. W. Erwin, S. -H. Lee and G. Shirane *Phys. Rev. B* **60**, 3643 (1999).
[38]   A. Bianconi, D. Di Castro, G. Bianconi, N. L. Saini, A. Pifferi, M. Colapietro F. C. Chou & D. C. Johnston *Physica C* Proc. of the M2S Houston Conf. 2000.
[39]   A. Bianconi, M. Missori, N. L. Saini, H. Oyanagi, H. Yamaguchi, D. H. Ha, Y. Nishihara *J. Superconductivity* **8,** 545 (1995); A. Bianconi, M. Missori, H. Oyanagi, H. Yamaguchi, D. H. Ha, Y. Nishihara, and S. Della Longa *Europhysics Lett*. **31**, 411 (1995);A. Bianconi, N. L. Saini, T. Rossetti, A. Lanzara, A. Perali, M. Missori, H. Oyanagi, H. Yamaguchi, and Y. Nishihara, D. H. Ha, *Phys. Rev. B* **54**, 12018 (1996); A. Bianconi, M. Lusignoli, N. L. Saini, P. Bordet, Å. Kvick, and P. Radaelli *Phys. Rev. B* **54**, 4310 (1996).
[40]   C. N. R. Rao and A. K. Ganguli *Chem. Soc. Rev*. **24**, 1 (1995).
[41]   P. P. Edwards, G. B. Peakok, J. P. Hodges, A. Asab, and I. Gameson in, *High $T_c$ Superconductivity: Ten years after the Discovery* (Nato ASI, *Vol*. 343) ed E. Kaldis, E. Liarokapis, and K. A. Müller, (Dordrecht, Kluwer) (1996) 135.
[42]   J. B. Goodenough *Supercond. Science and Technology* **3**, 26 (1990); J. B. Goodenough and A. Marthiram *J. Solid State Chemistry* **88**, 115 (1990)
[43]   S. Sachdev Quantum Phase Transitions (Cambridge Univ. Press, New York, 1999).